\documentclass[12pt, draftclsnofoot, onecolumn]{IEEEtran}
\usepackage{times,amsmath,color,amssymb,graphicx}
\usepackage{mathrsfs}
\usepackage{amsfonts}
\usepackage{amsthm}

\long\def\symbolfootnote[#1]#2{\begingroup%
\def\thefootnote{\fnsymbol{footnote}}\footnote[#1]{#2}\endgroup}

\begin{document}
\title{Riding on the Primary: A New Spectrum Sharing Paradigm for Wireless-Powered IoT Devices}
\author{~~~~~Xin Kang,~\IEEEmembership{Member,~IEEE},  ~Ying-Chang Liang,~\IEEEmembership{Fellow,~IEEE},
~Jing Yang,~\IEEEmembership{Member,~IEEE}
\thanks{X. Kang is with National Key Laboratory of Science and Technology on Communications, University of Electronic Science and Technology
of China, Chengdu, China 611731 (E-mail: kangxin@uestc.edu.cn). }
\thanks{Y. -C. Liang is with University of Electronic Science and Technology
of China, Chengdu, China 611731. He is also with School of Electrical and Information Engineering, The University of Sydney, NSW 2006, Australia. (E-Mail: liangyc@ieee.org). }
\thanks{J. Yang is with Singapore PowerGrid, Singapore Power Group,
2 Kallang Sector, Singapore 349277. (E-mail: jing@singaporepower.com.sg). }
\thanks{Part of this paper has been presented in \cite{KangBestPaper} at IEEE ICC 2017. }
}
\maketitle

\begin{abstract}
In this paper, a new spectrum sharing model referred to as riding on the primary (ROP) is proposed for wireless-powered IoT devices with ambient backscatter communication capabilities. The key idea of ROP is that the secondary transmitter harvests energy from the primary signal, then modulates its information bits to the primary signal, and reflects the modulated signal to the secondary receiver without violating the primary system's interference requirement. Compared with the conventional spectrum sharing model, the secondary system in the proposed ROP not only utilizes the spectrum of the primary system but also takes advantage of the primary signal to harvest energy and to carry its information. In this paper, we investigate the performance of such a spectrum sharing system under fading channels. To be specific, we maximize the ergodic capacity of the secondary system by jointly optimizing the transmit power of the primary signal and the reflection coefficient of the secondary ambient backscatter. Different (ideal/practical) energy consumption models, different (peak/average) transmit power constraints, different types (fixed/dynamically adjustable) reflection coefficient, different primary system's interference requirements (rate/outage) are considered.  Optimal power allocation and reflection coefficient are obtained for each scenario.
\end{abstract}


\section{Introduction}
Internet of Things (IoT) is a key application scenario of the fifth generation (5G) mobile communication systems.  It covers a wide range of use cases, such as smart home, smart wearables, smart farming, smart manufacturing, smart utilities, and smart city, which enable new business opportunities and new operational considerations for 5G. With the diverse use cases anticipated in IoT, the types of IoT devices are expected to diversified, and the characteristics and demands of different IoT devices are expected to vary a lot. Some of the devices, such as sensors and Radio-Frequency Identification (RFID) tags, are expected to be simple, small, low power, low throughput field devices. For this kind of IoT devices, a key requirement  from the industry  \cite{22861}  is that the power consumption should be very low and the battery life should be as long as ten years for extreme use cases. In these situations, energy harvesting, with potential to provide a perpetual power supply, becomes an attractive approach to prolong these devices' battery lifetime. Classic sources for energy
harvesting include solar and wind. Recently, ambient radio signal \cite{PowerwoWire}-\cite{KangXinEH} is receiving much research attention as a new viable source for energy harvesting, supported by the advantage that the wireless signals can carry both energy and information.

The backscatter communications technology used in RFID systems is a real-world application of energy harvesting from RF signals. \textcolor[rgb]{0.00,0.00,0.00}{In a typical backscatter system \cite{SRoyRFID,SRoyRFID2013}, the reader transmits a RF sinusoidal signal to a passive tag. The passive tag harvests RF energy from the signal to power its circuit, modulates its information bits onto the received sinusoidal signal by intentionally changing its amplitude and/or phase which is realized by changing its antenna impedance, and reflects the modulated signal back to the reader.} In \cite{VLiuBackscatter}, ambient backscatter, which is able to harvest energy from and transmits information over the ambient RF signals (e.g. TV signals), was proposed. In \cite{WiFiBackscatter}, Wi-Fi backscatter that uses the existing Wi-Fi infrastructure to provide internet connectivity for RF-powered devices was proposed. In \cite{FGaoBackscatter2016} and \cite{FGaoBackscatter}, maximum-likelihood detection was studied for an ambient backscatter system in which the tag adopts differential modulation. In \cite{GYangMITAir}, the modulator and the decoder design for backscattering over the ambient orthogonal frequency division multiplexing (OFDM) signals was studied. It was shown in recent work \cite{SKatti2015} and \cite{PZhang2016} that power harvested from ambient RF signals is sufficient to support the daily communication of a battery-less sensor through dedicated decoder and signal design. In \cite{KHuang2017}, it was further shown a network architecture with combined wireless power transfer and backscattering communication technology is able to a large and dense IoT network. However, these aforementioned works mainly focus on the hardware and decoder design but lacks fundamental system  analysis from theoretical aspects. Besides, the mutual influence (such as the interference) between the backscatter system and the primary system are not considered.

In this paper, we introduce the ambient backscatter communication technology to the cognitive radio (CR) system, and propose a new spectrum sharing model based on that. The proposed spectrum sharing model applies to CR systems with conventional primary communication systems and ambient-backscatter-based secondary systems. The key idea is that the secondary transmitter harvests energy from the primary signal, then modulates its information bits to the primary signal, and reflects the modulated signal to the secondary receiver without violating the primary system's interference requirement \cite{KangxinTWC}. The main contribution of this paper is summarized as follows:

\begin{itemize}
  \item We propose a new spectrum sharing model as Riding on the Primary (ROP). The differences between the proposed ROP and the existing technologies are as follows: (\textcolor[rgb]{0.00,0.00,0.00}{i) Compared with conventional backscatter communication systems, the reader (secondary receiver) in our system does not need to generate and transmit a RF sinusoidal signal, which can reduce its power consumption and prolongs its battery life. (ii) Compared with ambient backscatter communication systems, the interference from the wireless-powered tag (secondary transmitter) to the primary system is taken into consideration when designing the system. (iii) Compared with conventional spectrum sharing systems, the secondary system not only utilizes the spectrum of the primary system but also takes advantage of the primary system's signal transmission to carry its information.}
  \item We investigate the performance of the proposed ROP system under fading channels. To be specific, we maximize the ergodic capacity of the secondary system by jointly optimizing the transmit power of the primary signal and the reflection coefficient of the secondary ambient backscatter. Different (ideal/practical) energy consumption models, different (peak/average) transmit power constraints, different types (fixed/dynamically adjustable) reflection coefficient, different primary system's interference requirements (rate/outage) are considered.  Optimal power allocation and optimal reflection coefficient are obtained for each case. \textcolor[rgb]{0.00,0.00,0.00}{It is worth pointing out that certain degree of cooperation is needed between the primary and secondary system in order to do the joint optimization. For example, the primary system needs the knowledge of the secondary system (e.g., channel information) to optimize the transmit power, whereas the secondary system needs the knowledge of the primary system to optimize the reflection coefficient. }
  \item We show by numerical examples that dynamically adjustable reflection coefficient will result in a better system performance than a fixed reflection coefficient. The system performance under the average transmit power constraint is in general better than that under the peak transmit power constraint. Applying the primary transmission outage probability constraint to protect the primary system usually leads to a higher secondary ergodic capacity than adopting the primary transmission rate constraint. All these findings can serve as the guidance for designing high-performance practical ROP system.
\end{itemize}

The rest of this paper is organized as follows. Section II
introduces the system model of the proposed ROP spectrum sharing system.
Section III presents the basic problem formulation. Section IV investigates the capacity maximization
problem under the practical energy consumption model. Section V studies the capacity maximization
problem for fixed reflection coefficient and average transmit power constraint. Section VI considers the capacity maximization problem under the
primary transmission outage constraint. Then, numerical results are given in Section VII to verify the proposed studies. Finally, Section VIII concludes the paper.

\section{System Model}\label{Sec-systemmodel}

\begin{figure}
  \centering
  \includegraphics*[width=12cm]{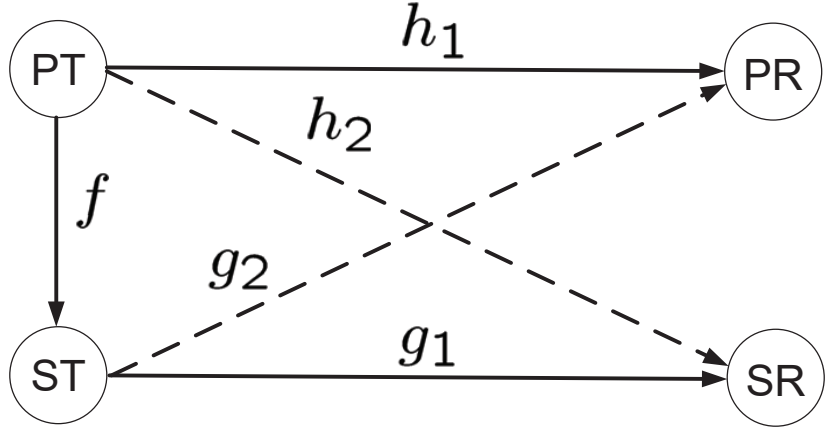}
\caption{System Model}\label{systemodel}
\end{figure}

\subsection{Channel Model}
In this paper, we consider a spectrum sharing communication system consists of a primary communication pair and a secondary transmission pair. The primary communication  pair is a conventional communication system consisting of a RF source (e.g. Base Stations, TV towers, WiFi APs) and a receiver (e.g., cell phones, TV receivers). The secondary communication pair is an ambient backscatter communication system which consists of a wireless-powered passive tag and a battery-powered reader.  For ease of explanation, we denote the RF source and the receiver  of the primary system as the primary transmitter (PT) and the primary receiver (PR), respectively. The wireless-powered tag and the reader of the backscatter system are denoted as secondary transmitter (ST) and secondary receiver (SR), respectively. In this paper, we consider the block fading channel model \cite{KangxinTWC}, where the channel coefficients remain the same for each block but may change from one block to another.  As shown in Fig. \ref{systemodel}, the channel power gains for the fading block $n$,  from the PT to the PR, from the ST to the SR are denoted by $h_1(n)$ and $g_1(n)$, respectively.  The channel power gains for cross channels for fading block $n$, i.e., from the PT to the ST, from the PT to the SR, and from the ST to the PR, are denoted by $f(n)$, $h_2(n)$ and $g_2(n)$, respectively.

\subsection{Transmission Model}
\textbf{Transmitted signal at the PT.}  Let $s(n;k)$ denote the transmitted signal of the PT at $k$th symbol of the $n$th block where $|s(n;k)|^2=1$,  and $p(n)$ denote the transmit power for fading block $n$. Then, the transmitted signal of the PT for the $k$th symbol of block $n$ is given by
\begin{align}x_{PT}(n;k)=\sqrt{p(n)}s(n;k), \forall k.\end{align}

\textbf{Transmitted signal at the ST.}  In fading block $n$, the signal received at the ST from the PT is $\sqrt{f(n)}x_{PT}(n;k)$. \textcolor[rgb]{0.00,0.00,0.00}{Note that the noise at the ST (Tag) is neglected as \cite{VLiuBackscatter,WiFiBackscatter,FGaoBackscatter2016,FGaoBackscatter,SKatti2015,BackNoise} since the on-tag integrated circuit only includes passive components.}  The power of the received signal at the ST from the PT is $f(n)p(n)$.  Part energy of the received signal is absorbed by the ST to power its circuit operation. The remaining part of the received signal is modified and backscattered to the reader. For convenience, we refer to this splitting factor as the reflection coefficient, and denote it by $\alpha(n)$ where $0\le \alpha(n)  \le 1$. Then, the energy of the transmitted signal of the ST can be denoted as $\alpha(n)f(n)p(n)$. Let $c(n;k)$ where $|c(n;k)|^2=1$ denote the ST's own signal, then the transmitted signal of the ST for the $k$th symbol of block $n$ is given by
\begin{align}x_{ST}(n;k)=\sqrt{\alpha(n)}\sqrt{f(n)}\sqrt{p(n)}s(n;k)c(n;k),\forall k.\end{align}
Note that we assume there is no signal processing delay of the backscatter circuit, i.e., there is no time delay between the transmitted signal and the received signal of the ST. This assumption is widely used in backscatter communication research literatures \cite{VLiuBackscatter}-\cite{GYangMITAir}.

\textbf{Received signal at the PR.}  Let $y_{PR}(n;k)$ denote the received signal at the PR for the $k$th symbol of block $n$, then we have \begin{align}y_{PR}(n;k)&=\sqrt{h_1(n)}x_{PT}(n;k)\nonumber\\&+\sqrt{g_2(n)}x_{ST}(n;k)+N_{PR}(n;k),\forall k,\end{align} where $N_{PR}(n;k)$ denotes the Gaussian receiving noise at the PR with zero mean and variance $\sigma_{PR}^2$. Then, the instantaneous received signal-to-interference-plus-noise ratio (SINR) at the PR for block $n$ denoted by $\gamma_{PR}(n)$ is given by
\begin{align}
\gamma_{PR}(n)=\frac{h_1(n) p(n) }{g_2(n) \alpha(n) f(n) p(n)+\sigma_{PR}^2}.
\end{align}

\textbf{Received signal at the SR.}   Let $y_{SR}(n;k)$ denote the received signal at the SR for the $k$th symbol of block $n$, then we have \begin{align}y_{SR}(n;k)&=\sqrt{g_1(n)}x_{ST}(n;k)\nonumber\\&+\sqrt{h_2(n)}x_{PT}(n;k)+N_{SR}(n;k),\forall k,\end{align} where $N_{SR}(n;k)$ denotes the Gaussian  receiving noise at the SR with zero mean and variance $\sigma_{SR}^2$. \textcolor[rgb]{0.00,0.00,0.00}{In this paper, we assume that SR decodes the received signal by performing successive interference cancellation (SIC). SIC is a well-known physical layer technique \cite{TCover,Verdu}. Briefly, SIC is the ability of a receiver to receive two or more signals concurrently. The SIC receiver decodes the stronger signal first, subtracts it from the combined signal, and extracts the weaker one from the residue. For the system setup considered in this paper, the secondary system is an ambient backscatter system. The strength of the signal received from the ST is in general much lower than that received from the PT (e.g., TV/WiFi signals). Thus, the SIC procedure at the SR is decoding the primary signal first and then subtracting it from the received signal before decoding its own signal.}  Thus, the
instantaneous received SNR at the SR for the block $n$ denoted by $\gamma_{SR}(n)$ is given by
\begin{align}
\gamma_{SR}(n)=\frac{g_1(n) \alpha(n) f(n) p(n)}{\sigma_{SR}^2}.
\end{align}

\textcolor[rgb]{0.00,0.00,0.00}{Note that we assume that the SR performs SIC decoding while the PR does not. This is due the fact that SIC needs to decode the stronger signal first. As mentioned above, the strength of the signal received from the ST is in general much lower than that received from the PT. Thus, SIC is not applicable at the PR. }

\section{Ergodic Capacity Maximization}\label{Sec-ErgodicCapacity}
Under the system model given in Section \ref{Sec-systemmodel}, the ergodic capacity of the secondary system can be written as
\begin{align}
\mathcal{C}_{SR}=\mathbb{E}\left[\log_2\left(1+\frac{g_1(n) \alpha(n) f(n) p(n)}{\sigma_{SR}^2}\right)\right],
\end{align}
where $\mathbb{E}\left[\cdot\right]$ denotes the statistic expectation, and it is taken over the joint fading states of the fading block $n$. In this paper, our objective is to maximize the ergodic capacity $\mathcal{C}_{SR}$ of the secondary system while guaranteing the performance of the primary system. In the following, we introduce the constraints that need to be considered when optimizing this network.

\textbf{PT's transmit power constraint. }
Let $P_{pk}$ denote the maximum transmit power of the PT, then the peak
transmit power constraint can be written as
\begin{align}\label{Con-peak}0\le p(n)\le P_{pk}, \forall n. \end{align}

\textbf{The reflection coefficient constraint.} Since the tag is a passive device, thus the energy harvested and reflected from the tag must be equal to the energy received from the primary signal. Thus, the reflection coefficient must satisfy the following constraint
\begin{align}\label{Con-refleco}0\le \alpha(n)\le 1, \forall n. \end{align}

\textbf{PR's rate constraint. } To guarantee the quality of service (QoS) of the primary system, we assume that there is a minimum rate requirement, which can be written as
\begin{align}\label{Con-rate}
\log_2\left(1+\frac{h_1(n) p(n) }{g_2(n) \alpha(n) f(n) p(n)+\sigma_{PR}^2}\right)\ge \gamma,\forall n.
\end{align}
where $\gamma$ is the minimum rate of the primary system.

\textbf{Tag's (ST's) circuit operation power requirement. } As aforementioned, the ST harvests energy from the primary signal to power its circuit operation.  Let $\epsilon_{ST}$ be the minimum power that the ST needs to support its circuit operation, then the following constraint must be satisfied in order for the ST to work, i.e.,
\begin{align}\label{Con-SustainPower}
\eta_{ST} (1-\alpha(n)) f(n) p(n)  \ge \epsilon_{ST},\forall n.
\end{align}
where $\eta_{ST}$ is the energy harvesting efficiency coefficient.

\textcolor[rgb]{0.00,0.00,0.00}{In this paper, the objective is to optimize the performance of such a spectrum sharing system by jointly optimizing the transmit power $p(n)$ of the PT and the reflection coefficient $\alpha(n)$ of the ST.}
The problems can be formulated as
\begin{align}
\underline{\textbf{P1:}}~~~~\underset{\{p(n),~\alpha(n)\}}{\mbox{Max}}~~&\mathcal{C}_{SR},\\
\mbox{s.t.}~~&\eqref{Con-peak},\eqref{Con-refleco},\eqref{Con-rate},\eqref{Con-SustainPower}.
\end{align}
For notation convenience, the fading block number $n$ is dropped from now on.

\textcolor[rgb]{0.00,0.00,0.00}{P1 is not a convex optimization problem. A problem is a convex optimization problem if its objective function is either convex or concave, and its feasible set is a convex set. For P1, its objective function is neither convex nor concave. This can be verified by looking at its Hessian matrix (A function is convex if its Hessian matrix is positive semi-definite). According to the composition rule given in \cite{ConvexOptimization}, for P1, we can determine the convexity of the objective function by investigating the convexity of $\alpha*p$. The Hessian matrix of $\alpha*p$ is $\left[0~1;~1~0\right]$, which is not positive semi-definite. Thus, P1 is not a convex optimization problem, and the conventional convex optimization techniques can not be applied to solve P1.} To solve P1, we first introduce the following Lemma.

\underline{\textbf{Lemma 1.}} The largest $\alpha$  that makes P1 feasible, denoted by $\alpha_L$, is given by
\begin{align}\label{alphaL}
\alpha_L= \max\kern-0.5mm\left\{\kern-0.5mm0, \kern-0.5mm\min\left\{\frac{1}{g_2 f}\kern-0.5mm\left(\frac{h_1}{2^\gamma\kern-0.5mm-\kern-0.5mm 1}\kern-0.5mm-\kern-0.5mm\frac{\sigma_{PR}^2}{ P_{pk}}\kern-0.5mm\right)\kern-0.5mm,1\kern-0.5mm-\kern-0.5mm\frac{\epsilon_{ST}}{\eta_{ST}fP_{pk}}\right\}\kern-0.5mm\right\},
\end{align}

\begin{proof}
The constraint \eqref{Con-rate} can be rewritten as
\begin{align}\label{Con-rate1}
\alpha \le \frac{h_1}{(2^\gamma-1)g_2 f }-\frac{\sigma_{PR}^2}{g_2 f p}.
\end{align}
It is observed that the right hand side of \eqref{Con-rate1} is an increasing function of $p$. Besides, due to the fact that $0\le p\le P_{pk}$,  the largest $\alpha$ denoted by $\alpha_{L1}$ satisfies \eqref{Con-rate1} is
\begin{align}
\alpha_{L1}=\frac{h_1}{(2^\gamma-1)g_2 f }-\frac{\sigma_{PR}^2}{g_2 f P_{pk}}.
\end{align}

Similarly, the constraint \eqref{Con-SustainPower} can be rewritten as
\begin{align}\label{Con-SustainPower1}
\alpha \le 1-\frac{\epsilon_{ST}}{\eta_{ST}pf}.
\end{align}
The right hand side of \eqref{Con-SustainPower1} is an increasing function of $p$.  Besides, due to the fact that $0\le p\le P_{pk}$,  the largest $\alpha$ denoted by $\alpha_{L2}$ satisfies \eqref{Con-SustainPower1} is
\begin{align}
\alpha_{L2} = 1-\frac{\epsilon_{ST}}{\eta_{ST}fP_{pk}}.
\end{align}

Thus, the largest $\alpha$  that satisfies both \eqref{Con-rate1} and \eqref{Con-SustainPower1} is given by $\min\{\alpha_{L1},\alpha_{L2}\}$.  Combining with the fact that $0\le \alpha  \le 1$, Lemma 1 follows.
\end{proof}

\underline{\textbf{Theorem 1.}} The optimal solution of P1 is given by
\begin{align}
p^*&=P_{pk},\\
\alpha^*&=\alpha_L,
\end{align}
where $\alpha_L$ is given by \eqref{alphaL}.
\begin{proof}
First, it is observed from P1 that all the constraints are instantaneous constraints. Thus, maximizing the ergodic capacity is equivalent to maximizing the instantaneous transmission rate, i.e., $\log_2\left(1+\frac{g_1 \alpha f p}{\sigma_{SR}^2}\right)$.   For any given $\alpha$ in the feasible region, the instantaneous rate is a monotonically increasing function with respect to $p$, and it attains the maximum value when $p=P_{pk}$. It is observed that for any given feasible $p$,  the instantaneous rate is a monotonically increasing function with respect to $\alpha$. Thus, $\alpha$ should be chosen as the largest $\alpha$ that makes P1 feasible, i.e., $\alpha_L$ given by \eqref{alphaL}, and it is shown in the proof of Lemma 1 that $\alpha_L$ is obtained when $p=P_{pk}$. Thus, it is clear that P1 is maximized when $p^*=P_{pk}$ and
$\alpha^*=\alpha_L$.
\end{proof}

\section{Practical Energy Consumption Model}
In this section, we consider a more practical energy consumption model of the ST's backscatter circuit, which is
\begin{align}\label{Con-PracticalEnergyModel}
\eta_{ST} (1-\alpha(n)) f(n) p(n)  \ge \epsilon_{b}+\epsilon_{s}\left(r_{ST}(n)\right),
\end{align}
where $\epsilon_{b}$ denotes the static energy consumption when the circuit is on, and  $\epsilon_{s}\left(r_{ST}(n)\right)$ denotes the dynamic energy consumption which is a function of its transmission rate. \textcolor[rgb]{0.00,0.00,0.00}{In practice, the dynamic energy consumption is in general proportional to the transmission rate. This is due to the following fact. The backscatter transmitter maps its bit sequence to RF waveforms by adjusting the load impedance of the antenna. The backscatter can control the rate at which it will generate the modulation symbols by controlling the switching frequencies on the SPDT (single pole double throw) switches \cite{SKatti2015}. In general, to achieve a higher data rate, the backscatter needs a higher frequency operation on the switches, which will cost more energy.} Thus, we model the dynamic energy consumption by $\epsilon_{s}\left(r_{ST}(n)\right)=u\log_2\left(1+\frac{g_1(n) \alpha(n) f(n) p(n)}{\sigma_{SR}^2}\right)$, where $u$ is a constant conversion parameter that relates the transmission rate with the energy consumption. In this section, under this energy consumption model, we re-investigate the optimization problem for this spectrum sharing system.

\begin{align}
\underline{\textbf{P2:}}~~~~\underset{\{p(n),~\alpha(n)\}}{\mbox{Max}}~~&\mathcal{C}_{SR},\\
\mbox{s.t.}~~&\eqref{Con-peak},\eqref{Con-refleco},\eqref{Con-rate},\eqref{Con-PracticalEnergyModel}.
\end{align}


The constraint \eqref{Con-PracticalEnergyModel} is a hyper-function with respect to $\alpha$ and $p$, which makes the problem difficult to solve. Thus, to solve P2, we first present the following two propositions.

\underline{\textbf{Proposition 1.}} Let $\hat{\alpha}$ denote the largest $\alpha$ that satisfies the constraint \eqref{Con-PracticalEnergyModel} for a given $p$, then $\hat{\alpha}$ can be obtained by solving the following equation:
\begin{align}\label{eq-CurveIntersect}
\eta_{ST} (1-\hat{\alpha}) f p=\epsilon_{b}+u\log_2\left(1+\frac{g_1 \hat{\alpha} f p}{\sigma_{SR}^2}\right).
\end{align}
\begin{proof}
It is observed that the left hand side of \eqref{Con-PracticalEnergyModel} (i.e., $\eta_{ST} (1-\alpha) f p$) is a monotonically decreasing function with respect to $\alpha$, while the right hand side of \eqref{Con-PracticalEnergyModel} ($\epsilon_{b}+u\log_2\left(1+\frac{g_1 \alpha f p}{\sigma_{SR}^2}\right)$) is a monotonically increasing function with respect to $\alpha$. Thus, it is easy to observe that the largest $\alpha$ is the intersection point of two curves, which is the solution of \eqref{eq-CurveIntersect}. Proposition 1 is thus proved.
\end{proof}

\underline{\textbf{Proposition 2.}} Let $\alpha_{B1}$ and $\alpha_{B2}$ be the solution of \eqref{eq-CurveIntersect} when $p=p_1$ and $p=p_2$, respectively. Then, we have
\begin{align}
\alpha_{B1}p_1 <  \alpha_{B2}p_2, ~\mbox{if}~p_1<p_2.
\end{align}

\begin{proof}
Since $\alpha_{B1}$ and $\alpha_{B2}$ be the solution of \eqref{eq-CurveIntersect} when $p=p_1$ and $p=p_2$, respectively. We have
\begin{align}\label{alpha_B1}
\eta_{ST} (1-\alpha_{B1}) f p_1=\epsilon_{b}+u\log_2\left(1+\frac{g_1 \alpha_{B1} f p_1}{\sigma_{SR}^2}\right).
\end{align}

\begin{align}\label{alpha_B2}
\eta_{ST} (1-\alpha_{B2}) f p_2=\epsilon_{b}+u\log_2\left(1+\frac{g_1 \alpha_{B2} f p_2}{\sigma_{SR}^2}\right).
\end{align}

Then, using \eqref{alpha_B2} to minus \eqref{alpha_B1}, we have
\begin{align}\label{eq-alphaVsP}
\eta_{ST}f (p_2\kern-0.5mm-\kern-0.5mmp_1\kern-1mm+\kern-0.5mm\alpha_{B1}p_1\kern-0.5mm-\kern-0.5mm\alpha_{B2}p_2)\kern-0.5mm=\kern-0.5mmu\log_2\kern-0.5mm\left(\kern-0.5mm\frac{g_1 f\alpha_{B2}  p_2\kern-0.5mm+\kern-0.5mm\sigma_{SR}^2}{g_1 f \alpha_{B1} p_1\kern-0.5mm+\kern-0.5mm\sigma_{SR}^2}\kern-0.5mm\right).
\end{align}

Then, in the following, we prove Proposition 2 by contradiction. Assume $\alpha_{B1}p_1 \ge \alpha_{B2}p_2$ when $p_1<p_2$. Then, under this presumption, it is clear that the left hand side of \eqref{eq-alphaVsP} is strictly positive, while the right hand side of \eqref{eq-alphaVsP} is zero or negative. This contradicts with the fact that the left hand side of \eqref{eq-alphaVsP} should be equal to the right hand side of \eqref{eq-alphaVsP}. Thus, our presumption does not hold. Thus, it follows that $\alpha_{B1}p_1 <  \alpha_{B2}p_2, ~\mbox{if}~p_1<p_2.$ Proposition 2 is thus proved.
\end{proof}

However, unlike P1, we cannot further prove $\alpha_{B1}<  \alpha_{B2}, ~\mbox{if}~p_1<p_2$. Thus, the approach used to solve P1 can not be applied here. Thus, to solve P2, we first consider the following problem,  which is

\begin{align}
\underline{\textbf{P2a:}}~\underset{\{p,~\alpha\}}{\mbox{Max}}~~&\mathcal{C}_{SR},\\
\mbox{s.t.}~~&\eqref{Con-peak},\eqref{Con-refleco},\eqref{Con-PracticalEnergyModel}.
\end{align}

Let $\alpha_{B}$ and $\alpha_{pk}$ be the solution of \eqref{eq-CurveIntersect} when $p=P_B$ and $p=P_{pk}$, respectively. Then, from Proposition 2, it follows
$\alpha_{B}P_B <  \alpha_{pk}P_{pk}, ~\forall P_B<P_{pk}.$
Since the objective function is an increasing function with respect to $\alpha p$, it is clear that the objective function attains its maximum value at $\alpha_{pk}P_{pk}$. Thus, the optimal solution of P2a can be obtained as
\begin{align}
p^*&=P_{pk},\\
\alpha^*&=\alpha_{pk},
\end{align}

Now, we return to P2. It is clear that the constraint \eqref{Con-rate} can be rewritten as
$\alpha \le \frac{h_1}{(2^\gamma-1)g_2 f }-\frac{\sigma_{PR}^2}{g_2 f p}.$
It is observed that its right hand side is an increasing function of $p$. Besides, due to the fact that $0\le p\le P_{pk}$,  the largest $\alpha$ satisfying $\alpha \le \frac{h_1}{(2^\gamma-1)g_2 f }-\frac{\sigma_{PR}^2}{g_2 f p}$ denoted by $\alpha_{M}$ is
\begin{align}\label{alphaM1}
\alpha_{M}=\frac{h_1}{(2^\gamma-1)g_2 f }-\frac{\sigma_{PR}^2}{g_2 f P_{pk}}.
\end{align}

Based on these results, we are now able to solve P2, and the solution is summarized in the following theorem.

\underline{\textbf{Theorem 2.}} The optimal solution of P2 is given by
\begin{align}
p^*&=P_{pk},\\
\alpha^*&=\min\left\{\alpha_{M},\alpha_{pk}\right\}
\end{align}
where $\alpha_{M}$ is given by \eqref{alphaM1}, and $\alpha_{pk}$ is the solution of \eqref{eq-CurveIntersect} when $p=P_{pk}$.

\begin{proof}
To prove Theorem 2, we consider the following two cases.

Case 1: $\alpha_{pk}\le \alpha_{M}$. In this case, the optimal solution of P2 is the same as that of P3a, which is $\alpha^*=\alpha_{pk}$ and $p^*=P_{pk}$.

Case 2: $\alpha_{pk}>\alpha_{M}$. In this case, the optimal $\alpha^*$ must satisfy the condition $0\le \alpha^*\le \alpha_{M}$. Now, we look at the following equation,
\begin{align}\label{eq-66}
\eta_{ST} (1-\alpha_{pk}) f P_{pk}=\epsilon_{b}+u\log_2\left(1+\frac{g_1 \alpha_{pk} f P_{pk}}{\sigma_{SR}^2}\right).
\end{align}
The above equation comes from the fact that $\alpha_{pk}$ is the solution of \eqref{eq-CurveIntersect} when $p=P_{pk}$.  Since $\alpha_{pk}>\alpha_{M}$, thus if we replace $\alpha_{pk}$ with $\alpha_M$, the left hand side of \eqref{eq-66} will increase, while the right hand side of \eqref{eq-66} will decrease. Thus, it follows
\begin{align}
\eta_{ST} (1\kern-0.5mm-\kern-0.5mm\alpha_{M}) f P_{pk}\ge \epsilon_{b}\kern-0.5mm+\kern-0.5mmu\log_2\left(1\kern-0.5mm+\kern-0.5mm\frac{g_1 \alpha_{M} f P_{pk}}{\sigma_{SR}^2}\right),
\end{align}
which indicates $\alpha_{M}$ and $P_{pk}$ is a feasible solution of P2.  Since $\alpha_{M}$ is the largest feasible $\alpha^*$ and $P_{pk}$ is the largest feasible $p^*$, thus the optimal solution is $\alpha^*=\alpha_{M}$ and $p^*=P_{pk}$.

Summarizing the above results, Theorem 2 follows.
\end{proof}

\section{Fixed Reflection Coefficient and Average Transmit Power Constraint}

\textbf{Fixed Reflection Coefficient. } In practice, for the purpose of circuit design simplicity, the reflection coefficient of the tag is designed to be fixed, i.e., the reflection cannot be dynamically changed in each fading block. To capture this fact, we introduce the following constraint
\begin{align}\label{Con-FixReflect}\alpha(n)=\alpha, \forall n.\end{align}

\textbf{Average transmit power constraint. } In practice, there is always a long-term power budget of the PT, and an average power constraint usually applies, which can be written as
\begin{align}\label{Con-APC}\mathbb{E}\left[p(n)\right]\le P_{av},\end{align}
where the statistic expectation is taken over the joint fading states of the fading block.

In the following subsections, we reinvestigate P1 and P2 under the above two practical constraints.
\subsection{Ideal Energy Consumption Model}
\begin{align}
\underline{\textbf{P3:}}~~~~\underset{\{p(n),~\alpha(n)\}}{\mbox{Max}}~~&\mathcal{C}_{SR},\\
\mbox{s.t.}~~&\eqref{Con-refleco},\eqref{Con-rate},\eqref{Con-SustainPower},\eqref{Con-FixReflect},\eqref{Con-APC}.
\end{align}


\textcolor[rgb]{0.00,0.00,0.00}{The objective function of P3 is the same as P1, and thus it is easy to verify that P3 is not a convex optimization problem. Therefore, it cannot be solved directly using convex optimization techniques. } Thus, for solving P3, we first consider P3 under a given $\alpha=\bar{\alpha}$. For given $\bar{\alpha}$, P3 can be rewritten as

\begin{align}
\underline{\textbf{P3a:}}~~~\underset{\{p(n)\ge 0\}}{\mbox{Max}}~~&\mathbb{E}\left[\log_2\left(1+\frac{g_1(n) \bar{\alpha} f(n) p(n)}{\sigma_{SR}^2}\right)\right],\\
\mbox{s.t.}~~&\mathbb{E}[p(n)] \le P_{av},\label{con-AV1}\\
&\log_2\left(1+\frac{h_1(n) p(n) }{g_2(n) \bar{\alpha} f(n) p(n)+\sigma_{PR}^2}\right)\ge \gamma,\label{con-AV2}\\
&\epsilon_{ST}-\eta_{ST} (1-\bar{\alpha}) f(n) p(n)  \le 0\label{con-AV3}.
\end{align}
Note the fading block number $n$ is dropped from now on for notation convenience.

P3a is a convex optimization problem since its objective function is concave and all the constraints are affine. To solve this problem, we first at the feasibility of the problem. Note that the constraint \eqref{con-AV2} is infeasible if $h_1-(2^\gamma-1)g_2 \bar{\alpha} f<0$, i.e., no matter how $p$ is chosen, \eqref{con-AV2} cannot be satisfied for such fading block. Thus, to save power, the optimal power allocation for such fading block is $p^*=0$.  When feasible,  it can be shown that constraints \eqref{con-AV2} and \eqref{con-AV3} can be rewritten as
$p\ge \frac{\sigma_{PR}^2(2^\gamma-1)}{h_1-(2^\gamma-1)g_2 \bar{\alpha} f}$ and $p\ge\frac{\epsilon_{ST}}{\eta_{ST} (1-\bar{\alpha}) f}$, respectively.   Thus, \eqref{con-AV2} and \eqref{con-AV3} can be replaced
by the following constraint
\begin{align}
p\ge P_m, \label{con-AV4}
\end{align}
where
\begin{align}\label{eq-Pm}
P_m\triangleq\max\left\{\frac{\sigma_{PR}^2(2^\gamma\kern-1mm-\kern-1mm1)}{h_1\kern-0.5mm-\kern-0.5mm(2^\gamma\kern-1mm-\kern-1mm1)g_2 \bar{\alpha} f},\frac{\epsilon_{ST}}{\eta_{ST} (1\kern-0.5mm-\kern-0.5mm\bar{\alpha}) f}\kern-0.5mm\right\}.
\end{align}

\underline{\textbf{Theorem 3.}} The optimal solution of P3a is given by
\begin{align}\label{Solution-T3}
p^*=\left\{\kern-1mm\begin{array}{ll}
0, & \mbox{if}~\lambda\kern-0.5mm\ge \frac{g_1 \bar{\alpha} f}{\sigma_{SR}^2\ln2},\\
             P_m, & \mbox{if}~\frac{g_1 \bar{\alpha} f}{\sigma_{SR}^2\ln2}>\kern-0.5mm\lambda\kern-0.5mm>\frac{g_1 \bar{\alpha} f}{\ln\kern-0.5mm2\left(\sigma_{SR}^2+g_1 \bar{\alpha} f P_m\right)},\\
             \frac{1}{\lambda\ln2}-\frac{\sigma_{SR}^2}{g_1 \bar{\alpha} f}, & \mbox{if}~\lambda\kern-0.5mm\le \frac{g_1 \bar{\alpha} f}{\ln\kern-0.5mm2\left(\sigma_{SR}^2+g_1 \bar{\alpha} f P_m\right)}.
           \end{array}
\right.
\end{align}
where $\lambda$ can be obtained by solving $\mathbb{E}\left[p^*\right]=P_{av}$, and $P_m$ is given by \eqref{eq-Pm}.
\begin{proof}
By introducing the dual variable associated with the average
transmit power constraint, the partial Lagrangian of P3a
problem is expressed as
\begin{align}
\mathcal{L}\left(p, \lambda\right)=\mathbb{E}\left[\log_2\left(1+\frac{g_1 \bar{\alpha} f p}{\sigma_{SR}^2}\right)\right]-\lambda\left(\mathbb{E}[p]-P_{av}\right),
\end{align}
where $\lambda$ is nonnegative Lagrange dual variable associated with constraints \eqref{con-AV1}.

Let $\mathcal{A}$ denote the set of $\left\{p\ge P_m\right\}$. The dual function is then expressed as
$q(\lambda) = \max_{p\in\mathcal{A}} ~\mathcal{L}\left(p, \lambda\right).$
The Lagrange dual problem is then defined as $\min_{\lambda\ge0} q(\lambda)$. The duality gap is zero for the convex
optimization problem addressed here, and thus solving its dual
problem is equivalent to solving the original problem.
\textcolor[rgb]{0.00,0.00,0.00}{The duality gap is zero if and only if strong duality holds. If the primal problem is convex and satisfies the Slater¡¯s condition \cite{ConvexOptimization}, then strong duality holds. The Slater's condition reduces to feasibility when the constraints are all linear equalities and inequalities  \cite{ConvexOptimization}. P3a is a convex problem and all its constraints are linear inequalities. It also can be verified that P3a is feasible under our constraints. Thus, the duality gap is zero for P3a.} Therefore,
according to the Karush-Kuhn-Tucker (KKT) conditions, the optimal solutions needs to satisfy the following
equations:
\begin{align}
\lambda^*\left(\mathbb{E}[p^*]-P_{av}\right)&=0,\label{eq-KKT1}\\
\lambda^* \ge0, ~~\mathbb{E}[p^*] -P_{av}&\le0,\label{eq-KKT1b}
\end{align}

For a fixed $\lambda$, by dual decomposition, the dual function
can be decomposed into a series of similar sub-dual-functions
each for one fading state. For a particular fading state, the
problem can be shown equivalent to
\begin{align}
\underline{\textbf{P3b:}}~~~~\underset{\{p\ge 0\}}{\mbox{Max}}~~&\log_2\left(1+\frac{g_1 \bar{\alpha} f p}{\sigma_{SR}^2}\right)-\lambda p,\\
\mbox{s.t.}~~&p\ge P_m,
\end{align}

It is easy to observe that the optimal solution of this subproblem is $p^*=+\infty$ if $\lambda=0$. Thus, in the following, we consider the optimal solution of this subproblem under the condition that $\lambda\neq0$.

The Lagrangian of this subproblem is
\begin{align}
\mathcal{L}_{sub}\left(p, \mu\right)&=\log_2\left(1+\frac{g_1 \bar{\alpha} f p}{\sigma_{SR}^2}\right)-\lambda p-\mu\left(P_m-p\right)+\nu p,
\end{align}
where $\mu$ and $\nu$ are nonnegative Lagrange dual variables associated with the constraints $p\ge P_m$ and  $p\ge 0$, respectively. Since the problem is convex, KKT conditions are sufficient to obtain its optimal solution. Thus, in the following, we investigate its KKT conditions:
\begin{align}
\frac{\partial \mathcal{L}\left(p^*, \mu^*\right)}{\partial p^*}\kern-0.5mm=\kern-0.5mm\frac{g_1 \bar{\alpha} f}{\ln\kern-0.5mm2\left(\sigma_{SR}^2+g_1 \bar{\alpha} f p^*\right)}\kern-0.5mm-\kern-0.5mm\lambda+\kern-0.5mm\mu^*+\kern-0.5mm\nu^*&=0,\label{eq-KKT2}\\
\mu^*\left(p^*-P_m\right)&=0,\label{eq-KKT3}\\
\nu^*p^*&=0,\label{eq-KKT5}\\
p^*-P_m&\ge 0,\label{eq-KKT4}\\
p^*\ge0,~\mu^*\ge0,~\nu^*&\ge0.
\end{align}
Now, we derive the optimal solution by solving these KKT conditions. To solve these KKT conditions, we consider the following two cases:

\begin{itemize}
  \item Case 1: $\mu^*>0$, $\nu^*=0$. For this case, it follows from \eqref{eq-KKT3} that \begin{align}\label{solution-sub1}p^*=P_m.\end{align} Then, based on \eqref{eq-KKT2}, $\mu^*$ can be obtained by solving $\frac{g_1 \bar{\alpha} f}{\ln\kern-0.5mm2\left(\sigma_{SR}^2+g_1 \bar{\alpha} f P_m\right)}\kern-0.5mm-\kern-0.5mm\lambda+\kern-0.5mm\mu^*=0$. Thus, \eqref{solution-sub1} holds only when $\lambda-\frac{g_1 \bar{\alpha} f}{\ln\kern-0.5mm2\left(\sigma_{SR}^2+g_1 \bar{\alpha} f P_m\right)}>0$.
  \item Case 2: $\mu^*>0$, $\nu^*>0$. In this case, it follows from \eqref{eq-KKT3} that $p^*=P_m$. However, it follows  from \eqref{eq-KKT5} that $p^*=0$. Thus, by contradictory, this case cannot happen.
  \item Case 3: $\mu^*=0$, $\nu^*=0$. For this case, it follows from \eqref{eq-KKT2} that
\begin{align}\label{solution-sub2}
p^*=\frac{1}{\lambda\ln2}-\frac{\sigma_{SR}^2}{g_1 \bar{\alpha} f}.
\end{align}
Then, taking \eqref{eq-KKT4} into consideration, \eqref{solution-sub2} holds only when $\frac{1}{\lambda\ln2}-\frac{\sigma_{SR}^2}{g_1 \bar{\alpha} f}\ge P_m$, i.e., $\lambda-\frac{g_1 \bar{\alpha} f}{\ln\kern-0.5mm2\left(\sigma_{SR}^2+g_1 \bar{\alpha} f P_m\right)}\le 0$.
  \item Case 4: $\mu^*=0$, $\nu^*>0$. It follows from \eqref{eq-KKT5} that \begin{align}\label{solution-sub2}p^*=0.\end{align} Then, based on \eqref{eq-KKT2}, $\nu^*$ can be obtained by solving $\frac{g_1 \bar{\alpha} f}{\sigma_{SR}^2\ln2}\kern-0.5mm-\kern-0.5mm\lambda+\kern-0.5mm\nu^*=0$. Thus, \eqref{solution-sub2} holds only when $\lambda-\frac{g_1 \bar{\alpha} f}{\sigma_{SR}^2\ln2}>0$.
\end{itemize}

Thus, combining the results obtained in Case 1 to 4,  the optimal solution for the subproblem can be summarized as \eqref{Solution-T3}.
Now, we have to find the optimal $\lambda^*$. As aforementioned, $\lambda^*$ has to satisfy \eqref{eq-KKT1} and \eqref{eq-KKT1b}.  It is observed that if $\lambda^*=0$, the optimal solution for the subproblem is
$p^*=+\infty$. This definitely violates the constraint $\mathbb{E}[p^*]\le P_{av}$. Thus, it follows that $\lambda^*\neq 0$. Then, according to the constraint $\lambda^*\left(\mathbb{E}[p^*]-P_{av}\right)=0$, it follows that the optimal $\lambda^*$ must satisfy $\mathbb{E}[p^*]=P_{av}$. Theorem 2 is thus proved.
\end{proof}

With the optimal solution of P3a given in Theorem 3, the optimal solution of P3 can be obtained by performing a one-dimension search for $\alpha$ over the space $\left[0,1\right]$.

\subsection{Practical Energy Consumption Model}
\begin{align}
\underline{\textbf{P4:}}~~~~\underset{\{p(n),~\alpha(n)\}}{\mbox{Max}}~~&\mathcal{C}_{SR},\\
\mbox{s.t.}~~&\eqref{Con-refleco},\eqref{Con-rate},\eqref{Con-PracticalEnergyModel},\eqref{Con-FixReflect},\eqref{Con-APC}.
\end{align}

\textcolor[rgb]{0.00,0.00,0.00}{The objective function of P4 is the same as P1, and thus it is easy to verify that P4 is not a convex optimization problem. Therefore, it cannot be solved directly using convex optimization techniques. } Thus, before solving P4, we first present the following proposition.
%

\underline{\textbf{Proposition 3.}} For a given $\bar{\alpha}$, the constraint \eqref{Con-PracticalEnergyModel} can be rewritten as
\begin{align}p\ge P_c,\end{align} where $P_c$ is the positive solution of $\eta_{ST} (1-\bar{\alpha}) f P_c  =\epsilon_{b}+u\log_2\left(1+\frac{g_1 \bar{\alpha} f P_c}{\sigma_{SR}^2}\right)$.
\begin{proof}
For the convenience of exposition, we introduce the following two functions $\mathcal{F}_1\left(p\right)\triangleq \eta_{ST} (1-\bar{\alpha}) f p$ and $\mathcal{F}_2\left(p\right)\triangleq \epsilon_{b}+u\log_2\left(1+\frac{g_1 \bar{\alpha} f p}{\sigma_{SR}^2}\right)$. It is easy to observe that both $\mathcal{F}_1\left(p\right)$ and $\mathcal{F}_2\left(p\right)$ are monotonically increasing functions with respect to $p$, and the increasing rate of $\mathcal{F}_1\left(p\right)$ is larger than that of $\mathcal{F}_2\left(p\right)$. It is also observed that $\mathcal{F}_1\left(0\right)\le \mathcal{F}_2\left(0\right)$. Thus, $\mathcal{F}_1\left(p\right)$ and $\mathcal{F}_2\left(p\right)$ must have a positive crossing point $P_c$. For any $p$ larger than $P_c$, we have $\eta_{ST} (1-\bar{\alpha}) f p \ge \epsilon_{b}+u\log_2\left(1+\frac{g_1 \bar{\alpha} f p}{\sigma_{SR}^2}\right)$. Proposition 3 is thus proved.
\end{proof}

It can be shown that constraints \eqref{Con-rate} can be rewritten as
$p\ge \frac{\sigma_{PR}^2(2^\gamma-1)}{h_1-(2^\gamma-1)g_2 \bar{\alpha} f}$. Thus, \eqref{Con-rate} and \eqref{Con-PracticalEnergyModel} can be replaced
by the following constraint
\begin{align}
p\ge P_L,
\end{align}
where
\begin{align}\label{eq-Pm2}
P_L\triangleq\max\left\{\frac{\sigma_{PR}^2(2^\gamma\kern-1mm-\kern-1mm1)}{h_1\kern-0.5mm-\kern-0.5mm(2^\gamma\kern-1mm-\kern-1mm1)g_2 \bar{\alpha} f},P_c\kern-0.5mm\right\}.
\end{align}

Using the same approach as P3, for a given $\bar{\alpha}$,  the optimal power allocation of P4 can be summarized as follows.

\underline{\textbf{Theorem 4.}} For a given $\bar{\alpha}$, the optimal power allocation of P4 is given by
\begin{align}
p^*=\left\{\kern-1mm\begin{array}{ll}
0, & \mbox{if}~\lambda\kern-0.5mm\ge \frac{g_1 \bar{\alpha} f}{\sigma_{SR}^2\ln2},\\
             P_L, & \mbox{if}~\frac{g_1 \bar{\alpha} f}{\sigma_{SR}^2\ln2}>\kern-0.5mm\lambda\kern-0.5mm>\frac{g_1 \bar{\alpha} f}{\ln\kern-0.5mm2\left(\sigma_{SR}^2+g_1 \bar{\alpha} f P_L\right)},\\
             \frac{1}{\lambda\ln2}-\frac{\sigma_{SR}^2}{g_1 \bar{\alpha} f}, & \mbox{if}~\lambda\kern-0.5mm\le \frac{g_1 \bar{\alpha} f}{\ln\kern-0.5mm2\left(\sigma_{SR}^2+g_1 \bar{\alpha} f P_L\right)}.
           \end{array}
\right.
\end{align}
where $\lambda$ can be obtained by solving $\mathbb{E}\left[p^*\right]=P_{av}$, and $P_L$ is given by \eqref{eq-Pm2}.

With the optimal solution of P4 under given $\bar{\alpha}$ obtained in Theorem 4, the optimal solution of P4 can be obtained by performing a one-dimension search for $\alpha$ over the the space $\left[0,1\right]$.

\section{Primary Transmission Outage Constraint}
In previous sections, we use the PR's rate constraint to guarantee the quality of service (QoS) of the primary system, i.e., there is a minimum rate requirement that the primary transmission has to fulfil. However, this constraint is too strict. In practice, certain ratio of transmission outage is usually acceptable. Thus, in this section, we introduce the \textbf{primary transmission outage constraint} \cite{KangxinJSAC2011}, which is mathematically defined as
\begin{align}\label{Eq-ProbRate}
\mbox{Prob}\left\{\log_2\left(1+\frac{h_1(n) p(n) }{g_2(n) \alpha(n) f(n) p(n)+\sigma_{PR}^2}\right)\le \gamma\right\}\le \epsilon,
\end{align}
where $\mbox{Prob}\left\{\cdot\right\}$ denotes the probability of the event, and $\epsilon$ denotes the maximum outage probability that is acceptable by the primary system.

\subsection{Peak Transmit Power Constraint}
We first consider the peak transmit power constraint case. Using the primary transmission outage constraint to replace the PR's rate constraint, and considering the fixed reflection coefficient, the problem can be formulated as
\begin{align}
\underline{\textbf{P5:}}~~~~\underset{\{p(n),~\alpha(n)\}}{\mbox{Max}}~~&\mathcal{C}_{SR},\\
\mbox{s.t.}~~&\eqref{Con-peak},\eqref{Con-refleco},\eqref{Con-SustainPower},\eqref{Con-FixReflect},\eqref{Eq-ProbRate}.
\end{align}

\textcolor[rgb]{0.00,0.00,0.00}{The objective function of P5 is the same as P1, and thus it is easy to verify that P5x is not a convex optimization problem. Therefore, it cannot be solved directly using convex optimization techniques. } We first consider P5 under a given $\bar{\alpha}$, i.e., $\alpha=\bar{\alpha}$ where $\bar{\alpha}$
is a constant.  For expression convenience, we denote the problem of P5 under a given $\bar{\alpha}$ as \textbf{P5a}.

\underline{\textbf{Theorem 5.}} The optimal power allocation of P5a, i.e., P5 under a given $\bar{\alpha}$, is given by
\begin{align}\label{Op-P5}
p^*(n)=P_{pk}, \forall n.
\end{align}


\begin{proof}
For a given $\bar{\alpha}$, the constraint \eqref{Con-SustainPower} can be rewritten as $p(n)\ge \frac{\epsilon_{ST}}{\eta_{ST}(1-\bar{\alpha})f(n)}$. For a given fading block $n$, it is easy to observe that P5a is infeasible when $\frac{\epsilon_{ST}}{\eta_{ST}(1-\bar{\alpha})f(n)}>P_{pk}$. This means that there does not exist a feasible $p(n)$ that can satisfy the ST's circuit operation power requirement, which indicates the secondary system does not work in this fading block. Thus, the contribution of this fading block to the secondary transmission's ergodic capacity is zero. The interference from the secondary system to the primary system is also zero. Thus, channel inversion power allocation or the maximum transmit power should be adopted for the primary system to minimize the possibility of outage for this fading block. Therefore, for simplicity, the optimal power allocation for such a fading block can be obtained as \begin{align}p^*(n)=P_{pk}.\end{align}

In the following, we consider the case that $\frac{\epsilon_{ST}}{\eta_{ST}(1-\bar{\alpha})f(n)}\le P_{pk}$.
To solve the problem, we introduce the following indicator function
\begin{align}
\chi\kern-1mm=\left\{\kern-1mm\begin{array}{ll}
              1, & \mbox{if}~\log_2\left(1+\frac{h_1(n) p(n) }{g_2(n) \alpha(n) f(n) p(n)+\sigma_{PR}^2}\right)\le \gamma,\\
              0, & \mbox{otherwsie}.
            \end{array}
\right.
\end{align}
For notation convenience, the fading block number $n$ is dropped from now on.
Then, it can be shown that the constraint given in \eqref{Eq-ProbRate} is equivalent to
\begin{align}\label{ExpectofChi}
\mathbb{E}\left[\chi\right]-\epsilon \le 0.
\end{align}

By introducing the dual variable associated with \eqref{ExpectofChi}, the partial Lagrangian of P5
under given $\bar{\alpha}$ can be expressed as
\begin{align}
\mathcal{L}\left(p, \lambda\right)=\mathbb{E}\left[\log_2\left(1+\frac{g_1 \bar{\alpha} f p}{\sigma_{SR}^2}\right)\right]-\lambda\left(\mathbb{E}[\chi]-\epsilon\right),
\end{align}
where $\lambda$ is nonnegative Lagrange dual variable associated with the constraint $\mathbb{E}\left[\chi\right]-\epsilon \le 0$.

Let $\mathcal{A}$ denote the set of $\left\{\frac{\epsilon_{ST}}{\eta_{ST}(1-\bar{\alpha})f} \le p\le P_{pk}\right\}$. The dual function is then expressed as
$q(\lambda) = \max_{p\in\mathcal{A}} ~\mathcal{L}\left(p, \lambda\right).$
According to the Karush-Kuhn-Tucker (KKT) conditions, the optimal solution needs to satisfy the following
equations:
\begin{align}
\lambda^*\left(\mathbb{E}[\chi^*]-\epsilon\right)&=0,\label{eq-KKT1}\\
\lambda^* \ge0, ~~\mathbb{E}[\chi^*] -\epsilon&\le0,\label{eq-KKT1b}
\end{align}

For a fixed $\lambda$, by dual decomposition, the dual function
can be decomposed into a series of similar sub-dual-functions
each for one fading state. For a particular fading state, the
problem can be shown equivalent to
\begin{align}
\underline{\textbf{P5b:}}~~~~\underset{\{p\ge 0\}}{\mbox{Max}}~~&\log_2\left(1+\frac{g_1 \bar{\alpha} f p}{\sigma_{SR}^2}\right)-\lambda \chi(p),\\
\mbox{s.t.}~~&p\ge \frac{\epsilon_{ST}}{\eta_{ST}(1-\bar{\alpha})f},\\
&p\le P_{pk}.
\end{align}

\begin{figure}
  \centering
  \includegraphics*[width=12cm]{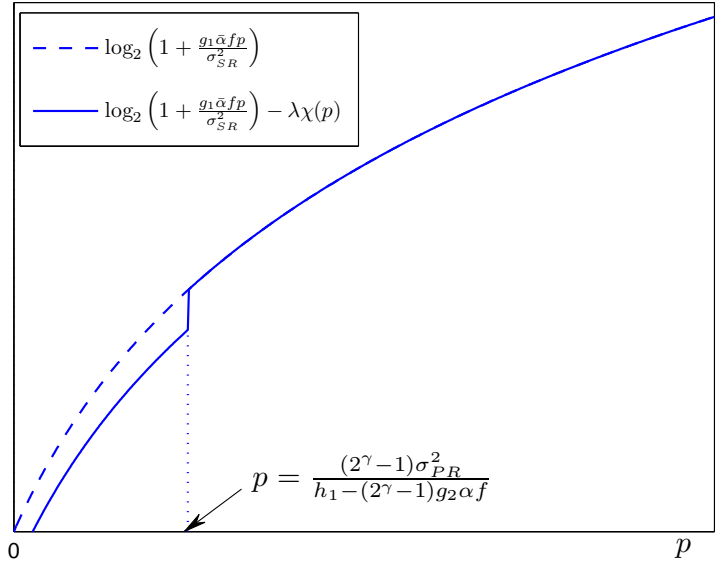}
\caption{An illustration of function $\log_2\left(1+\frac{g_1 \bar{\alpha} f p}{\sigma_{SR}^2}\right)-\lambda \chi(p)$.}\label{PPCFig1}
\end{figure}

Since $\log_2\left(1+\frac{g_1 \bar{\alpha} f p}{\sigma_{SR}^2}\right)$ is an increasing function, the optimal solution of P5b is $p^*=P_{pk}$ if $\lambda=0$.

In the following, we consider the optimal solution of this subproblem under the condition that $\lambda\neq0$. Fisrt, it can be observed that $\chi(p)$ can be rewritten as

\emph{Case 1:} $h_1-(2^\gamma-1)g_2\alpha f \le 0$.
             \begin{align}\label{chi_Form1}\chi(p)=1, \forall p.\end{align}

\emph{Case 2:} $h_1-(2^\gamma-1)g_2\alpha f >0$.
\begin{align}\label{chip}
\chi(p)=\left\{
\begin{array}{ll}
  1, & ~\mbox{if} ~p \le \frac{(2^\gamma-1)\sigma^2_{PR}}{h_1-(2^\gamma-1)g_2\alpha f},\\
  0, & ~\mbox{if} ~p >\frac{(2^\gamma-1)\sigma^2_{PR}}{h_1-(2^\gamma-1)g_2\alpha f}.
\end{array}
\right.
\end{align}

Let us first consider \emph{Case 1}. In this case, the objective function of P5b becomes $\log_2\left(1+\frac{g_1 \bar{\alpha} f p}{\sigma_{SR}^2}\right)-\lambda$, which is an increasing function with respect to $p$. Thus, the optimal solution of P5b is $p^*=P_{pk}$.

Now we consider \emph{Case 2}. It is observed from \eqref{chip} that $\chi(p)$ is a step function with respect to p for this case. The critical point is $p= \frac{(2^\gamma-1)\sigma^2_{PR}}{h_1-(2^\gamma-1)g_2\alpha f}$. Thus, the objective function of P5b has a form as illustrated in Fig. \ref{PPCFig1}. It can be observed from Fig. \ref{PPCFig1} that the function $\log_2\left(1+\frac{g_1 \bar{\alpha} f p}{\sigma_{SR}^2}\right)-\lambda \chi(p)$ is in general an increasing function of $p$. Thus, the optimal solution of P5b is $p^*=P_{pk}$.

Combining all these results, the optimal solution for P5a can be obtained as \eqref{Op-P5}. Theorem 5 is thus proved.
\end{proof}

With the optimal solution of P5a (i.e, P5 under given $\bar{\alpha}$) obtained in Theorem 5, the optimal solution of P5 can be obtained by performing a one-dimension search for $\alpha$ over the space $[0,1]$.

\subsection{Average Transmit Power Constraint}
Now, we investigate the problem under the average transmit power constraint, which is
\begin{align}
\underline{\textbf{P6:}}~~~~\underset{\{p(n),~\alpha(n)\}}{\mbox{Max}}~~&\mathcal{C}_{SR},\\
\mbox{s.t.}~~&\eqref{Con-refleco},\eqref{Con-SustainPower},\eqref{Con-FixReflect},\eqref{Con-APC},\eqref{Eq-ProbRate}.
\end{align}

We first consider P6 under a given $\bar{\alpha}$, i.e., $\alpha=\bar{\alpha}$ where $\bar{\alpha}$
is a constant.  For expression convenience, we denote the problem of P6 under a given $\bar{\alpha}$ as \textbf{P6a}.



Same as solving P5, it can be shown that \eqref{Eq-ProbRate} can be rewritten as \eqref{ExpectofChi}.
By introducing the dual variables associated with \eqref{ExpectofChi} and \eqref{Con-APC}, the partial Lagrangian of P6a
under given $\bar{\alpha}$ can be expressed as
\begin{align}
\mathcal{L}\left(p, \lambda\right)&=\mathbb{E}\left[\log_2\left(1+\frac{g_1 \bar{\alpha} f p}{\sigma_{SR}^2}\right)\right]-\lambda\left(\mathbb{E}[\chi]-\epsilon\right)\nonumber\\&-\mu\left(\mathbb{E}[p]-P_{av}\right),
\end{align}
where $\lambda$ and $\mu$ are nonnegative Lagrange dual variables associated with the constraints $\mathbb{E}\left[\chi\right]-\epsilon \le 0$ and $\mathbb{E}\left[\chi\right]-P_{av}\le 0$, respectively.

Let $\mathcal{A}$ denote the set of $\left\{\frac{\epsilon_{ST}}{\eta_{ST}(1-\bar{\alpha})f} \le p\right\}$. The dual function is then expressed as
$q(\lambda) = \max_{p\in\mathcal{A}} ~\mathcal{L}\left(p, \lambda\right).$
According to the Karush-Kuhn-Tucker (KKT) conditions, the optimal solution needs to satisfy the following
equations:
\begin{align}
\lambda^*\left(\mathbb{E}[\chi^*]-\epsilon\right)&=0,\label{eq-KKT1}\\
\mu^*\left(\mathbb{E}[p^*]-P_{av}\right)&=0,\\
\lambda^* \ge0, ~~\mathbb{E}[\chi^*] -\epsilon&\le0,\label{eq-KKT1b}\\
\mu^* \ge0, ~~\mathbb{E}[p^*] -P_{av}&\le0.
\end{align}

For a fixed $\lambda$, by dual decomposition, the dual function
can be decomposed into a series of similar sub-dual-functions
each for one fading state. For a particular fading state, the
problem can be shown equivalent to
\begin{align}
\underline{\textbf{P6b:}}~~~~\underset{\{p\ge 0\}}{\mbox{Max}}~~&\log_2\left(1+\frac{g_1 \bar{\alpha} f p}{\sigma_{SR}^2}\right)-\lambda \chi(p)-\mu p,\\
\mbox{s.t.}~~&p\ge \frac{\epsilon_{ST}}{\eta_{ST}(1-\bar{\alpha})f}.\label{con-P6a}
\end{align}
P6a can be solved by iteratively solving P6b for all fading states with fixed $\lambda$ and $\mu$, and updating these dual variables via sub-gradient based method, e.g., the ellipsoid method \cite{ConvexOptimization}, for which the details are omitted here for brevity.

In the following, we derive the solution to P6b. To solve the problem, we first introduce the following function
\begin{align}
\mathcal{F}(p)\triangleq\log_2\left(1+\frac{g_1 \bar{\alpha} f p}{\sigma_{SR}^2}\right)-\mu p.
\end{align}
It is easy to show that $\mathcal{F}(p)$ is a concave function with respect to $p$, and attains its maximum
value when $p$ is equal to
\begin{align}\label{p_tilde}
\tilde{p}=\frac{1}{\mu\ln 2}-\frac{\sigma_{SR}^2}{g_1\bar{\alpha}f}.
\end{align}

\begin{figure}
  \centering
  \includegraphics*[width=12cm]{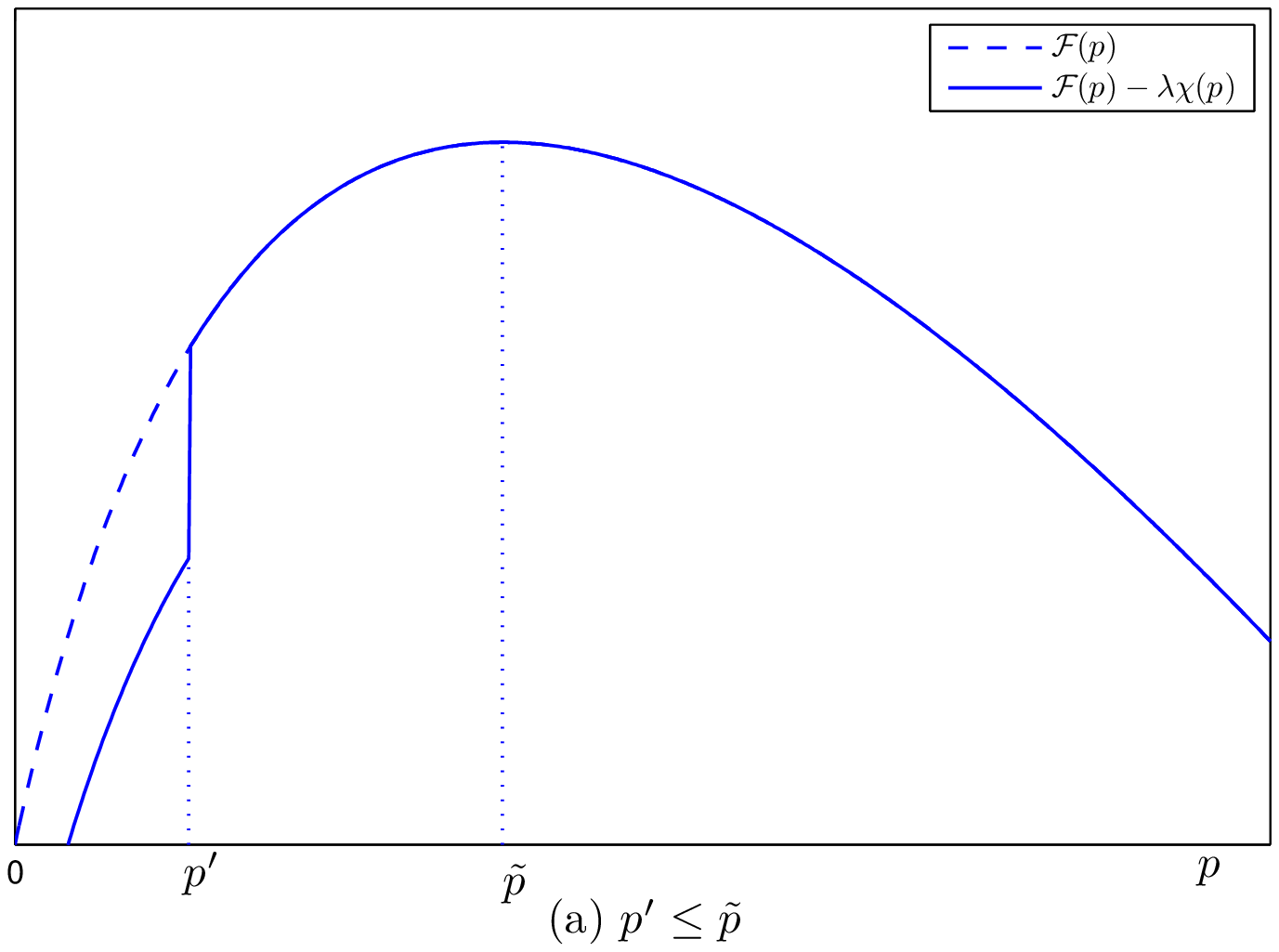}
   \includegraphics*[width=12cm]{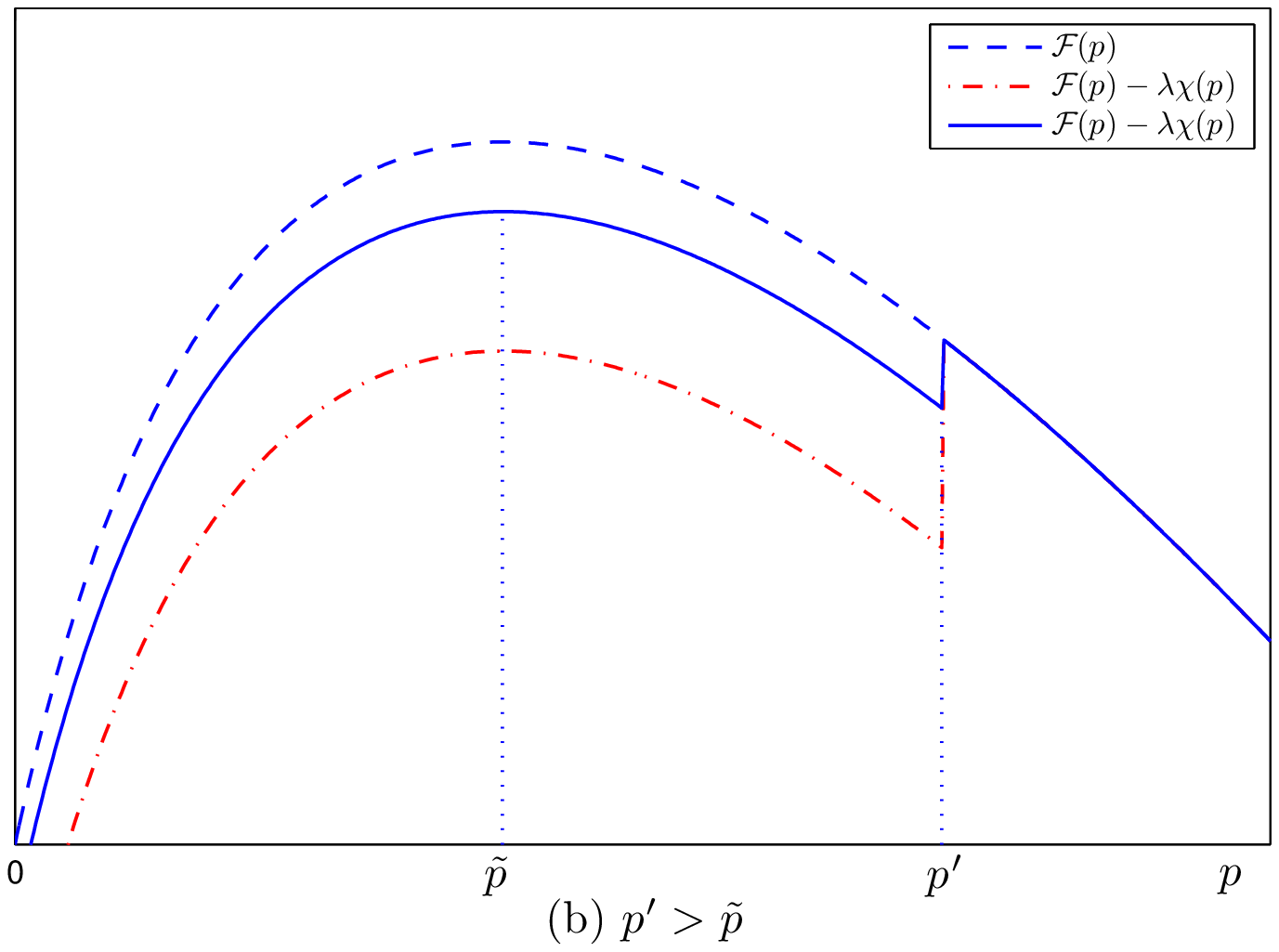}
\caption{An illustration of different forms of function $\mathcal{F}(p)-\lambda\chi(p)$.}\label{Fig1ac}
\end{figure}

Note that the objective function in P6b now becomes $\mathcal{F}(p)-\lambda \chi(p)$. Now we look at $\chi(p)$.
It has been shown in the previous subsection that $\chi(p)$ can be rewritten as \eqref{chi_Form1} and \eqref{chip}.

Thus, when $h_1-(2^\gamma-1)g_2\alpha f \le 0$,  we have $\chi(p)=1, \forall p$. Then, the objective function of P6b becomes $\mathcal{F}(p)-\lambda$, which attains its maximum value at the same $p$ as $\mathcal{F}(p)$. Thus, the optimal solution of P6b for this case is $p^*=\tilde{p}$.

When $h_1-(2^\gamma-1)g_2\alpha f > 0$, as shown in \eqref{chip}, $\chi(p)$ is a step function with respect to $p$, and its critical point is $p=\frac{(2^\gamma-1)\sigma^2_{PR}}{h_1-(2^\gamma-1)g_2\alpha f}$. In the following, we consider P6b for this case. For the convenience of discussion, we denote the critical point of  $\chi(p)$ as $p^{\prime}$, i.e.,
\begin{align}\label{p_prime} p^{\prime}=\frac{(2^\gamma-1)\sigma^2_{PR}}{h_1-(2^\gamma-1)g_2\alpha f}.\end{align}

For the convenience of discussion, we rewrite the constraint \eqref{con-P6a} as
$p\ge p^{\prime\prime}$, where
\begin{align}\label{p_primeprime}
p^{\prime\prime}=\frac{\epsilon_{ST}}{\eta_{ST}(1-\bar{\alpha})f}.
\end{align}

Let $p^*$ denote the optimal solution of P6b. The following discussions are then made on $p^*$:

\emph{Case 1}: $p^{\prime}\le \tilde{p}$. An illustration of this case is given in Fig. \ref{Fig1ac}(a).  It can be seen from Fig. \ref{Fig1ac}(a) that $\mathcal{F}(p)$ attains it maximum value at $p=\tilde{p}$, and at the value of $p=\tilde{p}$, $\chi(\tilde{p})$ is equal to zero. Thus, $\mathcal{F}(p)-\lambda\chi(p)$ attains its maximum value at $p=\tilde{p}$. Now, in order to obtain the optimal solution for P6a, we have to consider the following two subcases based on the relationship among $p^{\prime}$, $p^{\prime\prime}$, and $\tilde{p}$.
\begin{itemize}
  \item \emph{Subcase 1}: $p^{\prime\prime} \le \tilde{p}$. In this case, $\tilde{p}$ is within the feasible region. Thus, the optimal solution for P6b is $p^*=\tilde{p}$.
  \item \emph{Subcase 2}: $\tilde{p} < p^{\prime\prime}$. In this case, $\tilde{p}$ is not within the feasible region. It is observed from Fig. \ref{Fig1ac}(a) that $\mathcal{F}(p)-\lambda\chi(p)$ is a decreasing function with respect to $p$ for any $p\ge \tilde{p}$. Thus, the optimal solution for P6b is $p^*=p^{\prime\prime}$ for this case.
\end{itemize}

\emph{Case 2}: $p^{\prime}>\tilde{p}$. An illustration of this case is given in Fig. \ref{Fig1ac}(b). This case is a  little bit complex than \emph{Case 1}. As shown in Fig. \ref{Fig1ac}(b), $\mathcal{F}(p)-\lambda\chi(p)$ may attain its maximum value at $p=\tilde{p}$ or $p=p^{\prime}$ depending on the value of $\lambda$. Thus, the following three subcases are considered for finding the optimal $p^*$ for P6b.
\begin{itemize}
  \item \emph{Subcase 1}: $p^{\prime\prime} < \tilde{p}  < p^{\prime}$. For this case, both  $\tilde{p}$ and $p^{\prime}$ are within the feasible region. Thus, the optimal solution can be obtained as
      \begin{align}
      p^*=\left\{\begin{array}{ll}
                   \tilde{p}, & ~\mbox{if}~\mathcal{F}(\tilde{p})-\lambda>\mathcal{F}(p^{\prime}),  \\
                   p^{\prime}, & ~\mbox{if}~\mathcal{F}(\tilde{p})-\lambda\le \mathcal{F}(p^{\prime}).
                 \end{array}
      \right.
      \end{align}
  \item \emph{Subcase 2}: $\tilde{p} \le p^{\prime\prime} \le p^{\prime}$.  For this case, only $p^{\prime}$ is within the feasible region. However, it is possible that the value of $\mathcal{F}(p^{\prime\prime})-\lambda$ may be larger than that of $\mathcal{F}(p^{\prime})$. Thus, the optimal solution can be obtained as
            \begin{align}
      p^*=\left\{\begin{array}{ll}
                   p^{\prime\prime}, & ~\mbox{if}~\mathcal{F}(p^{\prime\prime})-\lambda>\mathcal{F}(p^{\prime}),  \\
                   p^{\prime}, & ~\mbox{if}~\mathcal{F}(p^{\prime\prime})-\lambda\le \mathcal{F}(p^{\prime}).
                 \end{array}
      \right.
      \end{align}
   \item \emph{Subcase 3}: $\tilde{p} < p^{\prime} < p^{\prime\prime}$. For this case, both  $\tilde{p}$ and $p^{\prime}$ are not within the feasible region.  It is observed from Fig. \ref{Fig1ac}(b) that $\mathcal{F}(p)-\lambda\chi(p)$ is a decreasing function with respect to $p$ for any $p\ge p^{\prime}$. Thus, the optimal solution for P6b is $p^*=p^{\prime\prime}$ for this case.
\end{itemize}

Combining all the results obtained above, the optimal solution for P6a can be summarized in the following theorem.

\underline{\textbf{Theorem 6.}} The optimal power allocation of P6a, i.e., P6 under a given $\bar{\alpha}$, is given by
\begin{align}
p^*(n)=\left\{
\begin{array}{cc}
  \tilde{p}, & ~\mbox{if}~\mathbb{R}_1~\mbox{holds},\\
  p^{\prime},& ~\mbox{if}~\mathbb{R}_2~\mbox{holds}, \\
  p^{\prime\prime},& ~\mbox{if}~\mathbb{R}_3~\mbox{holds},
\end{array}
\right.
\end{align}
where $\tilde{p}$, $p^{\prime}$, $p^{\prime\prime}$ are given by \eqref{p_tilde}, \eqref{p_prime}, \eqref{p_primeprime}, respectively. $\mathbb{R}_1$, $\mathbb{R}_2$, $\mathbb{R}_3$ are defined as $\mathbb{R}_1\triangleq\left\{\left\{h_1-(2^\gamma-1)g_2\alpha f \le 0\right\}\right.$ or $\left\{h_1-(2^\gamma-1)g_2\alpha f > 0\right.$
and $\left.\tilde{p}>\max\left\{p^{\prime},p^{\prime\prime}\right\}\right\}$ or $\left\{p^{\prime}>\tilde{p}>p^{\prime\prime}\right.$ and $h_1-(2^\gamma-1)g_2\alpha f > 0$ and  $\left.\left.\mathcal{F}(\tilde{p})-\lambda>\mathcal{F}(p^{\prime})\right\}\right\}$.

$\mathbb{R}_2\triangleq$ $\left\{\left\{h_1-(2^\gamma-1)g_2\alpha f > 0\right.\right.$
and $p^{\prime}>\tilde{p}>p^{\prime\prime}$ and $\left.\mathcal{F}(\tilde{p})-\lambda<\mathcal{F}(p^{\prime})\right\}$ or $\left\{p^{\prime}>p^{\prime\prime}>\tilde{p}\right.$ and
$h_1-(2^\gamma-1)g_2\alpha f > 0$
and  $\left.\left.\mathcal{F}(p^{\prime\prime})-\lambda<\mathcal{F}(p^{\prime})\right\}\right\}$.

$\mathbb{R}_3\triangleq$ $\left\{\left\{h_1-(2^\gamma-1)g_2\alpha f > 0\right.\right.$
and $\left.p^{\prime\prime}>\max\left\{p^{\prime},\tilde{p}\right\}\right\}$ or
$\left\{h_1-(2^\gamma-1)g_2\alpha f > 0\right.$
and $p^{\prime}>p^{\prime\prime}>\tilde{p}$ and $\left.\left.\mathcal{F}(p^{\prime\prime})-\lambda>\mathcal{F}(p^{\prime})\right\}\right\}$.

With the optimal solution of P6a (i.e, P6 under given $\bar{\alpha}$) obtained in Theorem 6, the optimal solution of P6 can be obtained by performing a one-dimension search for $\alpha$ over the space $[0,1]$.


\section{Numerical Results}
In this section, several numerical examples are presented to evaluate the performance of the
derived results. 
\textcolor[rgb]{0.00,0.00,0.00}{We assume i.i.d. Rayleigh fading for all channels
involved, and thus the channel power gains of these channels are exponentially distributed. The stochastic mean of the channel power gain is assumed to be one. It is worth pointing
out that the assumption of particular distributions of the channel power gains does not affect the
structure of the problem studied and the solution obtained.} The power of the noises at the receiver of PR and SR are assumed to be one. The energy harvest efficiency $\eta_{ST}$ for ST is assumed to be one. The constant conversion parameter $u$ is assumed to be one. The results given in these examples are obtained by averaging over $10000$ channel realizations.

\subsection{Ideal Vs. Practical Energy Consumption Model}
In this subsection, we investigate the impact of the energy consumption model on the system performance for two different scenarios given below.
\begin{figure}
  \centering
  \includegraphics*[width=12cm]{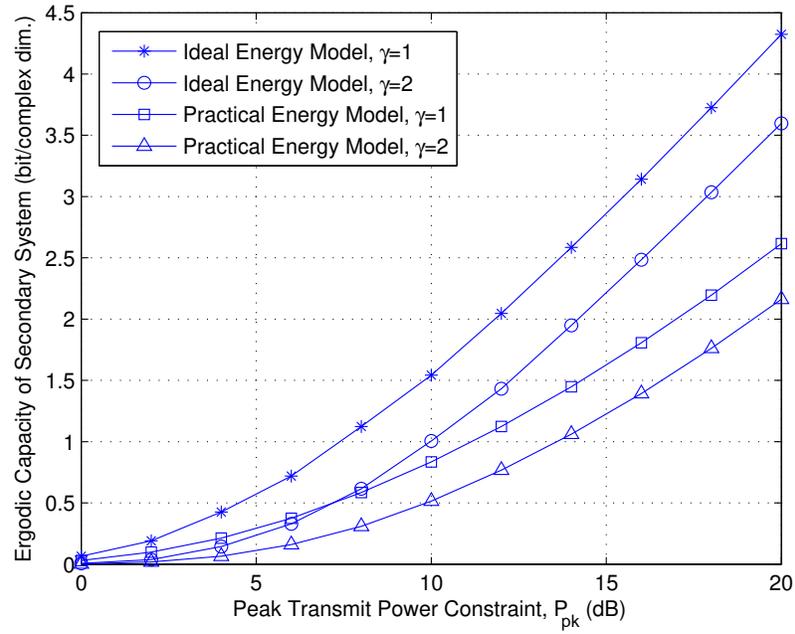}
\caption{Peak Transmit Power Constraint with Dynamic Reflection Coefficient}\label{Fig1}
\end{figure}

\begin{figure}
  \centering
  \includegraphics*[width=12cm]{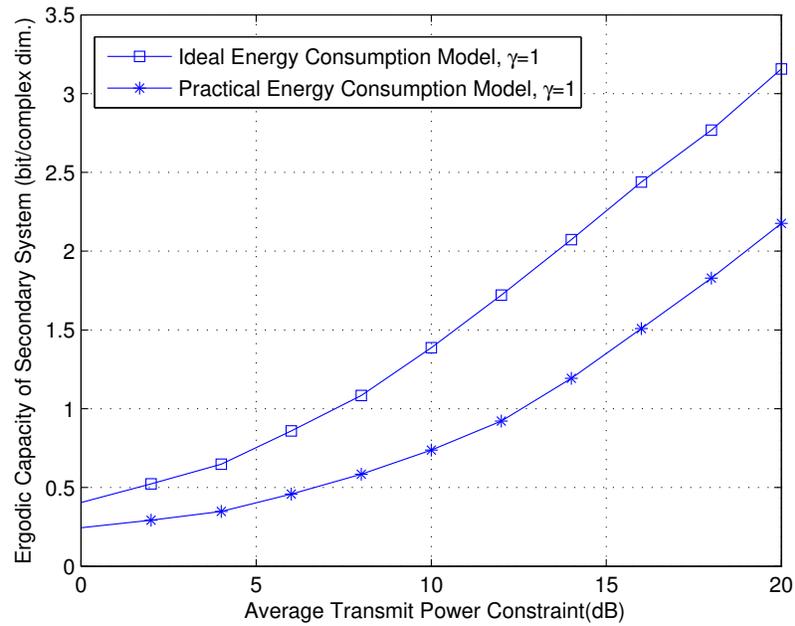}
\caption{Average Transmit Power Constraint with Fixed Reflection Coefficient}\label{Fig2}
\end{figure}

\subsubsection{Peak transmit power constraint with dynamic reflection coefficient} \textcolor[rgb]{0.00,0.00,0.00}{The curves presented in Fig \ref{Fig1} are obtained based on the solutions to P1 and P2.} It is observed from Fig. \ref{Fig1} that the ergodic capacity of the secondary system increases with the increasing of $P_{pk}$ for all curves. It is also observed that the ergodic capacity of the ideal energy consumption model is larger than that of the practical energy model for the same $P_{pk}$. This is due to the fact that for the practical energy consumption model, the dynamic power consumption is considered, more power is needed to support the tag's circuit operation, and thus the power left for transmitting the signal is less, which results in a lower transmission rate. We also observe that the capacity for $\gamma=1$ is larger than that for $\gamma=2$ for both idea and practical energy consumption models. This is as expected since a lower primary system's rate requirement indicates a high interference tolerance, and thus the secondary system can transmit with a higher power which results in a higher transmission rate.

\subsubsection{Average transmit power constraint with fixed reflection coefficient} In Fig.\ref{Fig2}, we show that ergodic capacity for both ideal and practical energy consumption model under the  average transmit power constraint and the fixed reflection coefficient. \textcolor[rgb]{0.00,0.00,0.00}{The curves presented in Fig \ref{Fig2} are obtained based on the solutions to P3 and P4.} The trend of the curves are same  as that of Fig.\ref{Fig1}. Thus, for concise, the explanations are not repeated here. However, comparing Fig.\ref{Fig2} with Fig.\ref{Fig1}, we observe that the ergodic capacity in Fig.\ref{Fig2} is lower than that of Fig.\ref{Fig1} when $P_{pk}=P_{av}$ for the same energy consumption model and the same $\gamma$. This is due to the fact the reflection coefficient in Fig.\ref{Fig1} can be dynamically adjusted to its optimal value for each fading block while the reflection coefficient in Fig.\ref{Fig2} remains the same for each fading block.

\subsection{Study of the Primary Transmission Outage Constraint}
In this subsection, we investigate the impact of the primary transmission outage constraint on the system performance for two different setups given below.

\subsubsection{Peak transmit power constraint} 
\begin{figure}
  \centering
  \includegraphics*[width=12cm]{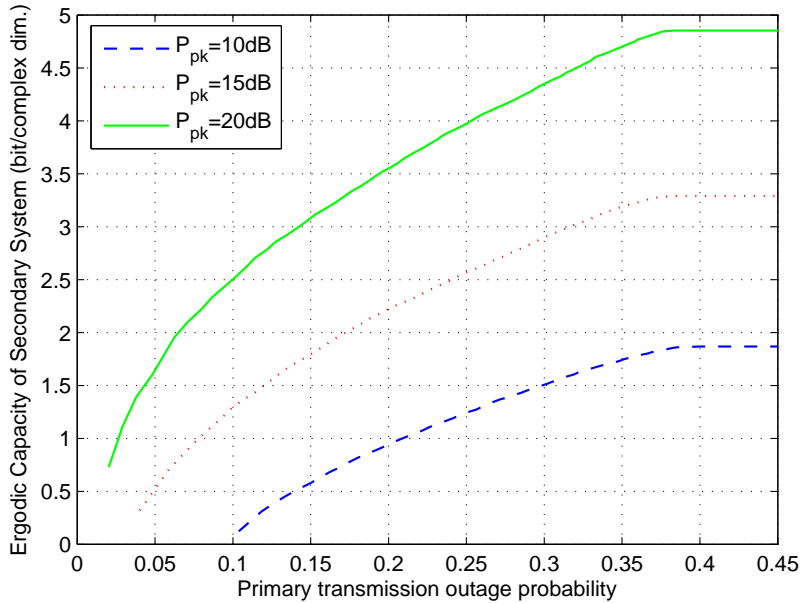}
\caption{Peak Power Constraint with Primary Transmission Outage Constraint}\label{Fig3}
\end{figure}

\textcolor[rgb]{0.00,0.00,0.00}{The curves presented in Fig \ref{Fig3} are obtained based on the solution to P5.} Firstly, it is trivial to observe from Fig. \ref{Fig3} that the ergodic capacity under larger $P_{pk}$ is larger for the same given primary transmission outage probability. Secondly, the ergodic capacities for all three curves become flat when the primary transmission outage probability is sufficiently large. This is as expected since the peak transmit power constraint becomes the bottleneck that limits the system performance when the primary transmission outage probability is large. It is also observed that all three curves are not starting from the zero primary transmission outage probability. This is because the zero primary transmission outage probability is in fact equivalent to the PR's rate constraint given in \eqref{Con-rate}. For a fixed $\alpha$ and a given peak power constraint $P_{pk}$, there does not exist a power allocation that satisfies $\log_2\left(1+\frac{h_1(n) p(n) }{g_2(n) \alpha f(n) p(n)+\sigma_{PR}^2}\right)\ge \gamma$ for a given positive $\gamma$ for all the fading slots due to the continuous distribution of the channel power gains.

\begin{figure}
  \centering
  \includegraphics*[width=12cm]{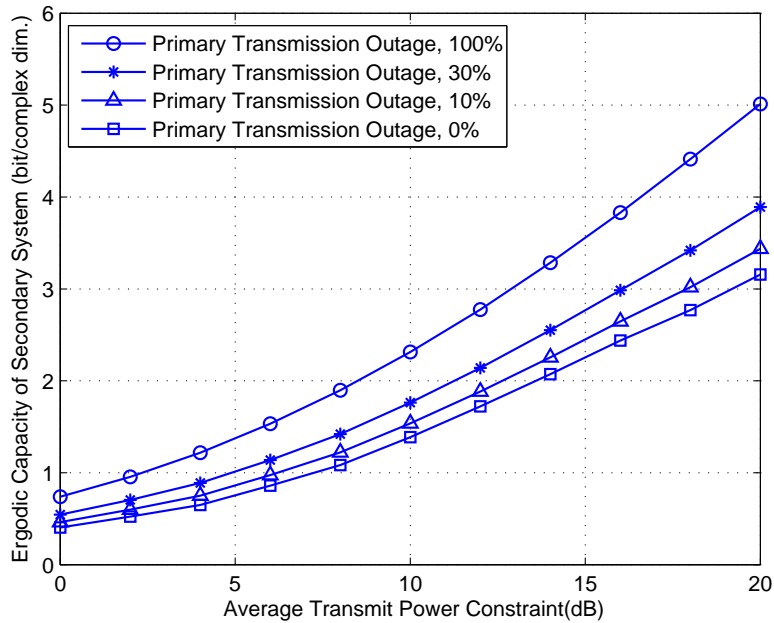}
\caption{Average Power Constraint with Primary Transmission Outage Constraint}\label{Fig4}
\end{figure}

\subsubsection{Average transmit power constraint} In Fig. \ref{Fig4}, four curves with different primary transmission outage probability are given.  \textcolor[rgb]{0.00,0.00,0.00}{The curves presented in Fig \ref{Fig4} are obtained based on the solution to P6.} It is easy to observe that the ergodic capacities increase with the increasing of the transmit power constraint. The capacity difference for four curves is small when the transmit power constraint is small. This is intuitive since the average transmit power constraint is the bottleneck that limits the system performance when it is very small. It is also observed that the ergodic capacity under larger primary outage probability is larger for the same given transmit power constraint. Besides, the curves with $100\%$ and $0\%$ primary outage probability serve as the upper bound and the lower bound, respectively. This is as expected since the case under $100\%$ primary outage probability is equivalent to the case without the primary outage probability constraint. The case under $0\%$ primary outage probability is equivalent to the case under the primary rate constraint which is studied in P3.

\section{Conclusions}
In this paper, we proposed the Riding on the Primary (ROP) spectrum sharing model for wireless-powered IoT devices with ambient backscatter communication capabilities. We investigated the performance of such a spectrum sharing system under fading channels. The ergodic capacity of the secondary system was investigated by jointly optimizing the transmit power of the primary signal and the reflection coefficient of the secondary system. Different (ideal/practical) energy consumption models, different (peak/average) transmit power constraints, different types (fixed/dynamically adjustable) reflection coefficient, different primary system's interference requirements (rate/outage) are considered were considered. Closed-form solutions were obtained for most cases. Performance for different scenarios were studied and compared through numerical simulations.

\end{document}